\documentclass[hyper]{JHEP} 

\usepackage{epsfig}

\newcommand\fverb{\setbox\pippobox=\hbox\bgroup\verb}
\newcommand\fverbdo{\egroup\medskip\noindent%
			\fbox{\unhbox\pippobox}\ }
\newcommand\fverbit{\egroup\item[\fbox{\unhbox\pippobox}]}
\newbox\pippobox

\title{Proposal for non-BPS D-brane action}

\author{by J. Kluso\v{n}\\
	 Department of Theoretical Physics and Astrophysics\\
                   Faculty of Science, Masaryk University\\
Kotl\'{a}\v{r}sk\'{a} 2, 611 37, Brno\\
Czech Republic\\
	E-mail: \email{klu@physics.muni.cz}}

\preprint{\hepth{0004106}}

\abstract{In this short note we would like to propose a form for the
action of a non-BPS D-brane, which will be manifestly supersymmetric 
invariant and T-duality covariant. We also explicitly show that tachyon
condensation on the world-volume of this brane leads to the Dirac-Born-Infeld 
action for BPS D(p-1)-brane.}


\begin{document}

\section{Introduction}

In the paper \cite{Sen} Sen proposed a supersymmetric invariant
Dirac-Born-Infeld (DBI) action for a non-BPS D-brane.  Since the
non-BPS branes break all supersymmetries, it seams to be strange to
construct a supersymmetric action describing this brane. However,
although there is no manifest supersymmetry of the world-volume theory,
we still expect the world-volume theory to be supersymmetric, with the
supersymmetry realised as a spontaneously broken symmetry.  {}From
these arguments Sen showed that the action has to contain the full
number of fermionic zero modes (=32), because they are fermionic
Goldstone modes of the completely broken supersymmetry, while a BPS D-brane
contains 16 zero modes, because it breaks one half of the supersymmetry. Sen
showed that the DBI action for the non-BPS D-brane (without the presence
of the tachyon) is the same as the supersymmetric action describing the
BPS D-brane.  This action is manifestly invariant under all space-time
supersymmetries. Sen argued that the ordinary action for the BPS
D-brane contains the DBI term and the Wess-Zumino (WZ) term, which are
invariant under supersymmetry. But only when they both are present in
the action for the D-brane, the action is invariant under the local $
\kappa$-symmetry, which is needed for gauging away one half of the
fermionic degrees of freedom, so that only 16 physical fermionic
fields remain on the BPS D-brane, as should be the case for an object breaking
16 bulk supersymmetries.  Sen showed that the DBI term for a non-BPS
D-brane is exactly the same as the DBI term in the action of a BPS
D-brane (when we suppose that other massive fields are integrated out,
including the tachyon) that is invariant under supersymmetry
transformations, but has no $ \kappa$-symmetry, so that the number of
fermionic degrees of freedom is 32 which is the appropriate number of
fermionic Goldstone modes for an object braking bulk
supersymmetry completely.

Sen also showed how we could include the tachyonic field into the
 action.  Sen proposed a form of the term expressing the interaction
 between the tachyon and other light fields on the world-volume of a
 non-BPS D-brane on the grounds of invariance under the
 supersymmetry and general covariance.  This term has the useful
 property that for a constant tachyon field it is zero, so that the
 action for the non-BPS D-brane in that case vanishes identically.

There have been many attempts to generalise Sen's proposal for the construction
of non-BPS D-branes. In a previous paper \cite{Kluson} we tried to
construct the action for a non-BPS Dp-brane in Type IIA theory, which
was manifestly supersymmetric and also had the property that tachyon
condensation in the 
form of a kink solution leads to the BPS D-brane in Type IIA theory.
Different forms of the action for a non-BPS D-brane was suggested in
\cite{Berkof1,Garrousi}. These proposals reflects the remarkable symmetry
between the tachyon and other massless degrees of freedom, which was
anticipated in \cite{Horava}. The action presented in \cite{Berkof1}
was T-duality covariant, while the action presented in \cite{Kluson}
was not T-duality covariant, which is the main problem of that
proposal.

In the present paper we would like to propose yet another form of the
action for a single non-BPS D-brane which will combine the virtues of
all previous attempts.  We propose the form of the action, which will
be manifestly 
supersymmetry invariant, T-duality covariant and in the linear
approximation the equation of motion for the tachyon will naturally
have a smooth tachyon kink solution as a solution. As a result, the action
for a non-BPS Dp-brane will reduce to the Dirac-Born-Infeld action for
D(p-1)-brane, which together with the tachyon condensation in the
Wess-Zumino (WZ) term for a non-BPS D-brane \cite{Billo,Berkof1} will
lead to the supersymmetric action for a BPS D(p-1)-brane.

We will also discuss the tachyon kink solution for the 
action proposed in \cite{Berkof1,Garrousi}. We will find a
remarkable fact that the tachyon kink solution in the form
of a piecewise function is a natural solution of the linearised equation
of motion obtained from this action regardless to the form
of the tachyon potential. 

In the conclusions, we will suggest possible extensions of this work.
\section{Proposal  for   non-BPS D-brane action}
We start this section with recapitulating the basic facts
 about non-BPS D-branes in Type IIA theory, following \cite{Sen}
\footnote{For non-BPS D-brane the situation is basically the
 same with difference in
chirality of the Majorana-Weyl fermions. 
We refer to \cite{Sen,Aganagic} for more details.}.
Let $ \sigma^{\mu} , \ \mu=0,...p $ are 
world-volume coordinates on a D-brane.
 Fields living
on this D-brane arise as the lightest states from 
the spectrum of the open string ending on this D-brane. 
These open strings have two CP sectors \cite{Sen2}:
 The first one, with unit $ 2\times 2 $ matrix, which corresponds
to the states of the open string with the usual GSO projection
 $ (-1)^{F}\left| \psi \right>=\left| \psi \right > $, 
 where $ F$ is the world-sheet fermion number and 
$ \left| \psi \right> $ is a state from the Hilbert space of
the open string living on a Dp-brane. The second CP sector
 has CP matrix  $ \sigma_1 $ and contains states
having opposite GSO projection 
$ (-1)^{F} \left| \psi \right>=-\left| \psi \right> $. 
The massless fields living
on a Dp-brane are ten components of $ X^M(\sigma), M=0,...,9 $ ;  
a $ U(1) $ gauge field $ A(\sigma)_{\mu} $ and
 a fermionic field $ \theta $ with $32$ real components 
transforming as a Majorana spinor under the Lorenz
group $ SO(9,1) $. We can write $ \theta $ as a sum of a 
left-handed Majorana-Weyl spinor and a right-handed
Majorana-Weyl spinor:
\begin{equation}
\theta=\theta_L+\theta_R ,\ \Gamma_{11}\theta_L=\theta_L, \
\Gamma_{11}\theta_R=-\theta_R 
\end{equation}
All fields except $ \theta_R $ come from the CP sector with the identity
matrix, while $\theta_R $ comes from  
the sector with the  $ \sigma_1 $ matrix 
\footnote{Our conventions are following. $\Gamma^M$ are $32\times 32$
Dirac matrices 
appropriate to 10d with the relation $\{\Gamma^M,\Gamma^N\}=2\eta^{MN}$, with 
$\eta^{MN}=(-1,1,...,1)$. For this choice of gamma matrices the
massive Dirac equation 
is $(\Gamma^M\partial_M-M)\Psi=0$. We also introduce $\Gamma_{11}=
\Gamma_0...\Gamma_9 , (\Gamma_{11})^2=1$.}.

As Sen \cite{Sen} argued, the action for a non-BPS D-brane 
(without tachyon) should go to the action for BPS D-brane, when
we set $ \theta_R=0 $ (we have opposite convention that \cite{Sen}). 
For this reason, the action
for a non-BPS D-brane in \cite{Sen} was constructed 
as the supersymmetric DBI action but without $ \kappa$-symmetry so
that we are not able to gauge away half of the fermionic 
degrees of freedom. This action thus describes a non-BPS D-brane.

The next thing is to include is the effect of the tachyon.
 In order to get some relation between tachyon condensation and
the supersymmetric D-branes, we would like to have an effective action
 for the massless fields and the tachyon living on the world-volume of  
a non-BPS D-brane. Following \cite{Sen}, the effective action
 for a  non-BPS Dp-brane with a tachyonic field on its
world-volume should has a form:
\begin{equation}\label{action}
 S=-C_p\int d^{p+1}\sigma\sqrt{-\det
(\mathcal{G}_{\mu\nu}+2\pi\alpha'\mathcal{F} 
_{\mu\nu} )}F(T,\partial T, \theta_L,\theta_R,\mathcal{G},..), 
\end{equation}

\begin{equation}
\Pi^M_{\mu}=\partial_{\mu}X^M-\overline{\theta}
\Gamma^M\partial_{\mu}\theta \ , \ \mathcal{G}_{\mu\nu}=
\eta_{MN}\Pi^M_{\mu}\Pi^N_{\nu} 
\end{equation}
and
\begin{equation}
\mathcal{F}_{\mu\nu}=F_{\mu\nu}-[\overline{\theta}
\Gamma_{11}\Gamma_M\partial_{\mu}\theta 
(\partial_{\nu}X^M-\frac{1}{2}\overline{\theta}\Gamma^M
\partial_{\nu}\theta) -(\mu \leftrightarrow \nu ) ].
\end{equation}
The constant  
$ C_p=\sqrt{2}T_p=\frac{2\pi\sqrt{2}}{g(4\pi^2\alpha')^{\frac{p+1}{2}}} $ is
a tension for  a non-BPS Dp-brane, where $T_p$ is a tension for  a BPS Dp-brane
and $g$ is a string coupling constant. The function $F$
  contains the dependence of  the tachyon and its 
derivatives and  may also depend on  other world-volume
and background field.

We have proposed \cite{Kluson}
an action for a non-BPS D-brane in Type IIA theory in the form
\begin{equation}\label{actionKluson}
 S=-C_p\int d^{p+1}\sigma\sqrt{-\det
(\mathcal{G}_{\mu\nu}+2\pi\alpha'\mathcal{F} 
_{\mu\nu} )}F(T,\partial T, \theta_L,\theta_R,\mathcal{G},..), 
\end{equation}
where the function $F$ takes the form
\footnote{The meaning of $\tilde{\mathcal{G}}^{\mu\nu}_S$
will be explained latter.}
\begin{equation}\label{F}
 F=\left(
\tilde{\mathcal{G}}_S^{\mu\nu}\partial_{\mu}T\partial_{\nu}T
+V(T)+I_{TF}\right), 
\end{equation}
where $I_{FT}$ contains
interaction terms between the tachyon and fermionic fields, which
was determined on the base of the supersymmetric invariance. However,
this term also contains the expression
 $f(T)\tilde{\mathcal{G}}^{\mu\nu}_S\partial_{\mu}
   \overline{\theta}_R\partial_{\nu}\theta_L $,
which, as was shown in \cite{Berkof1}, is not T-duality covariant,
so that this term should not be present in the action.  On the other
hand, the equation of motion for the tachyon obtained from (\ref{actionKluson})
do lead to the tachyon kink solution and the non-BPS D-brane 
reduces to the BPS D-brane of codimension one and the presence of the 
term cited above leads to the elimination of one half of fermionic
degrees of freedom,
which suggests that the resulting kink solution is a BPS D-brane. Then we 
 argued that through tachyon condensation we have restored
the $\kappa$-symmetry on the world-volume of the resulting D-brane. It
can seam that the elimination of the term $f(T)\tilde{\mathcal{G}}^{\mu\nu}_S
\partial_{\mu}\overline{\theta}_R\partial_{\mu}\theta_L$ on the grounds
of T-duality covariance will lead to conclusion that through tachyon
condensation 
we are not able to obtain BPS D-brane. However, as we will show, this is
not completely true. We must also say that the term $I_{TF}$ contain many
interaction terms with a difficult structure, while the interaction between 
fermions and tachyon presented in \cite{Berkof1} emerges in a very natural and
symmetric way. This seems to tell us to follow their approach in the 
construction of the action for a non-BPS D-brane.

In this paper we would like to propose 
the Dirac-Born-Infeld action for a single non-BPS
D-brane in the form
\begin{equation}\label{Naction}
S_{DBI}=-C_p\int d^{p+1}\sigma V(T)
\sqrt{ -\det\left(\mathcal{G}_{\mu\nu}+
2\pi\alpha'\mathcal{F}_{\mu\nu}+
\frac{2\pi\alpha'\partial_{\mu}T\partial_{\nu}T}{V(T)}\right)},
\end{equation}
where $V(T)$ is a tachyonic potential, which in the zeroth 
order approximation is equal to \cite{SenT,Berkovits}
\begin{equation}\label{zeropot}
V(T)=-2\pi\alpha'm^2T^2/2+\lambda T^4+\frac{(2\pi\alpha'm^2)^2}{16\lambda}=
\lambda (T^2-T_0^2)^2,
\end{equation}
where $m^2=\frac{1}{2\alpha'}, T_0^2=\frac{2\pi\alpha'm^2}{4\lambda}$. 
In the following we do not need to know  the explicit value of the
constant $\lambda$.

We must stress that the form of the  action (\ref{Naction}) was mainly
inspired with the recent proposals \cite{Berkof1,Garrousi} where
the action for non-BPS D-brane was given as
\begin{equation}\label{actionBerkof}
S=-C_p\int d^{p+1}\sigma V(T)
\sqrt{-\det (\mathcal{G}_{\mu\nu}+2\pi\alpha'\mathcal{F}_{\mu\nu}
+2\pi\alpha'\partial_{\mu}T\partial_{\nu}T)} \ .
\end{equation}
We have modified the action given above to the action
(\ref{Naction}) in order to get smooth tachyon kink solution
\cite{SenT}. Than we will show that the tachyon condensation
in the action (\ref{Naction}) leads naturally to the DBI action for
BPS  D-brane and
together with the tachyon condensation in
 the Wess-Zumino term for a non-BPS D-brane proposed
in \cite{Billo} and generalised to the supersymmetric invariant form
in \cite{Berkof1}
\begin{equation}\label{WZ}
S_{WZ}=\int C \wedge dT \wedge e^{ 2\pi
\alpha' \mathcal{F}},
\end{equation}
gives a correct description of a non-BPS D-brane in the approximation
of slowly varying fields. However we will see in the 
next section that the equation of motion for the tachyon
obtained from the linearised form of the action (\ref{actionBerkof})
leads to solution which is a piecewise tachyon kink solution
regardless the form of the tachyon potential. 

In this paragraph we will discuss the properties of the 
action (\ref{Naction}). The action is manifestly supersymmetric
invariant, since contains the supersymmetric invariant terms
\cite{Sen,Aganagic}, together with a natural requirement that
the tachyon field
is invariant under supersymmetric transformations.  The action
is  manifestly invariant under the world-volume reparametrisation as well.
The action is also T-duality covariant \cite{Simon2}. This can be
easily seen from the fact that the potential $V(T)$ does not change
under T-duality transformation and the T-duality covariance of the
other terms in (\ref{Naction}) was proven in \cite{Berkof1,Simon2}.

The action (\ref{Naction}) is equal to zero for the tachyon equal to its
vacuum value $T_0$ which can be seen from  the fact that for $T=const$
its derivative is equal to zero while $V(T)\neq 0$ for $T\neq T_0$, so that
the action reduces to the action  anticipated by Sen \cite{Sen} for the
case of a constant tachyon field
\begin{equation}
S=-C_p\int d^{p+1}\sigma V(T)\sqrt{-\det (
\mathcal{G}_{\mu\nu}+2\pi\alpha'\mathcal{F}_{\mu\nu})}
\rightarrow 0 \ ,\mathrm{if} \  \ T\rightarrow T_0.
\end{equation}
We can also see that for $T=0$, which corresponds to the $(-1)^{F_L}$
operation \cite{SenA} that takes a non-BPS D-brane in Type
IIA (IIB) theory into a BPS D-brane in Type IIB(IIA) theory, the derivative
of the tachyon is zero, so that the term $\sim \partial T$ in the action
(\ref{Naction}) is equal to zero, while $V(0)$ is nonzero and depends
on the precise form of the tachyon potential. For example, for the zeroth
order approximation of the  tachyon potential,  $V(T=0)$ is equal
about $0.60$ of 
the tension of a non-BPS D-brane \cite{SenT} so that the DBI action for
a non-BPS D-brane in Type IIA (IIB) theory goes to the DBI action for
a BPS D-brane  
in Type IIB(IIA) theory (of course, with appropriate modification of 
fermionic terms, since in Type IIB theory we have spinors of the same chirality)
with the tension
\begin{equation}
T_p= 0.6T_p^c,
\end{equation}
where $T_p^c=\frac{2\pi}{g(4\pi^2\alpha')^{(p+1)/2}}$ is the
correct tension for a BPS Dp-brane. In the previous equation
we have used the transformation rule for the tension of the
non-BPS D-brane under $(-1)^{F_L}$ operation \cite{SenA}
: $(-1)^{F_L}: C_p\rightarrow T_p$. We believe that with the inclusion
of the higher order corrections to the tachyon potential we get the
exact result. 

It is also easy to see that the action given in (\ref{Naction}) reduces
into the action (\ref{actionBerkof}), when we neglect the higher powers
of the tachyon field  in the expression
$2\pi\alpha'\partial_{\mu}T\partial_{\nu}TV(T)^{-1}$, 
because the tachyon potential must contain the constant term ensuring
that the potential is equal to zero for the tachyon equal to its
vacuum value. Then we  
have
\begin{equation}
2\pi\alpha'\partial_{\mu}T\partial_{\nu}T V(T)^{-1}
\approx A\cdot2\pi \alpha'\partial_{\mu}T\partial_{\nu}T
+ O(T^{4}),
\end{equation}
where $A$ is some constant that depends on the precise form
of the tachyon potential. As was argued in \cite{Simon2}, the requirement
of T-duality do not precisely fix the numerical constant in front of
the term $(\partial T)^2$, so that the presence of constant $A$ do not
affect the similarity with the term given in (\ref{actionBerkof}).

As a last check we will show that in the linear approximation
the action (\ref{Naction}) reduces to the action (\ref{actionKluson})
without the interaction terms $I_{TF}$ between the fermions and the tachyon. 
In fact,
the interaction between tachyon and fermions is included directly
in the form of the DBI action, which can be easily seen from the rewriting
the determinant in (\ref{Naction}) in the form
\begin{equation}\label{deter}
\det \left[(\mathcal{G}+2\pi\alpha'\mathcal{F})_{\mu\nu}\right]
 \det \left[\delta^{\mu}_{\nu}
+2\pi\alpha'\tilde{\mathcal{G}}_S^{\mu\kappa}\partial_{\kappa}T
\partial_{\nu}TV(T)^{-1}\right ],
\end{equation}
where $\tilde{\mathcal{G}}^{\mu\nu}$ is a inverse of $\mathcal{G}+\mathcal{F}$
and $ (...)_S$ means the symmetric part of a given matrix. This result follows
from the fact that $\partial_{\mu}T\partial_{\nu}T$ is symmetric in
the world-volume 
indexes.
When we expand the second determinant in (\ref{deter}) and when we restrict
ourselves to the  linear approximation, then we obtain from (\ref{Naction})
\begin{equation}\label{actlinear}
S_{DBI}=-C_p\int d^{p+1}\sigma \sqrt{-\det (\mathcal{G}_{\mu\nu}+
2\pi\alpha'\mathcal{F}_{\mu\nu})} \left(V(T)+\frac{2\pi\alpha'}{2}
\tilde{\mathcal{G}}_S^{\mu\nu}\partial_{\mu}T\partial_{\nu}T\right),
\end{equation}
which is the same action as (\ref{actionKluson}) without the fermionic
terms. 

We can show that the equation of motion for the tachyon obtained from 
(\ref{actlinear}) leads naturally to the solution, which has the behaviour
of a kink solution. This solution has been given earlier in \cite{Kluson} and
we will review this calculation.

We get the equation of motion for the tachyon from the variation
of (\ref{actlinear}), which give (we consider dependence of the
tachyon on only one of the coordinates, $x$ say):
\begin{equation}\label{eq}
D\left(\frac{d}{dx}\left(\frac{\delta F}{\delta \partial_x T}\right)
-\frac{d F}{dT}\right)+
(2\pi\alpha')\partial_{\mu}D\tilde{\mathcal{G}}_S^{\mu x}\partial_{x}T=0,
\end{equation}
where $F$ has a form:
\begin{equation}\label{F2}
F=\left(
\frac{2\pi\alpha'}{2}\tilde{\mathcal{G}}_S^{\mu\nu}
  \partial_{\mu}T\partial_{\nu}T
+V(T)\right) 
\end{equation}
and we have defined
\begin{equation}
D=\sqrt{-\det (\mathcal{G}_{\mu\nu}+(2\pi\alpha')\mathcal{F}_{\mu\nu})}.
\end{equation}

Firstly we will consider the first bracket in (\ref{eq}). 
The first expression in (\ref{eq}) gives
\begin{equation}\label{kt1}
2\pi\alpha'\partial_{\mu}(\tilde{\mathcal{G}}^{\mu x}_S\partial_x T(x) )
=2\pi\alpha'\partial_{\mu}(\tilde{\mathcal{G}}_S^{\mu x})\partial_x T+
2\pi\alpha'\tilde{\mathcal{G}}_S^{x x}\partial_x\left(\partial_x T\right)
,\end{equation}
where we have used the fact that the tachyon field is a function of
$x$ only. Since 
for tachyon in the form of a kink solution the first derivative is
nonzero, the  first term 
in (\ref{kt1}) leads to the  result
\begin{equation}\label{con}
\tilde{\mathcal{G}}^{\mu x}_S=const.
\end{equation}

Since the constant in (\ref{con}) has not any physical meaning we 
can take solution in the form
\begin{eqnarray}\label{Gcon}
\tilde{\mathcal{G}}^{\mu x}_S=0 , x\neq \mu , \
\tilde{\mathcal{G}}^{xx}_S=1  \ \Rightarrow \nonumber \\
G_{xx}=1 ,G_{x\mu }=0, \nonumber \\
\end{eqnarray}
where $G_{\mu\nu}$ is a inverse matrix
of $\tilde{\mathcal{G}}_S^{\mu\nu}$ and plays the role of
 the natural open string metric \cite{Witen2}.

With using (\ref{Gcon}), the second expression in (\ref{eq}) gives the condition 
\begin{equation}\label{FGcon}
\partial_x D=0 \Rightarrow
\partial_x \mathcal{G}_{\mu\nu}=\partial_x
\mathcal{F}_{\mu\nu}=0.
\end{equation}
When we return to (\ref{Gcon}) and use 
 \begin{equation}
G_{\mu\nu}=\mathcal{G}_{\mu\nu}-(2\pi\alpha')^2
\mathcal{F}_{\mu\kappa}\mathcal{G}^{\kappa\delta}
\mathcal{F}_{\delta\nu},
\end{equation}
we get
\begin{equation}
1=\mathcal{G}_{xx}-(2\pi\alpha')^2\mathcal{F}_{x\alpha}\mathcal{G}^
{\alpha\beta}\mathcal{F}_{\beta x},
\end{equation}
where $\alpha,\beta=0,...,p-1 , x=x^p$.
Since $\mathcal{G}^{\alpha\beta}\neq 0$, we obtain the natural
solution of the previous equation in the form:
\begin{equation}\label{kt3}
\mathcal{G}_{xx}=1 ,\mathcal{G}_{x\alpha}=
\mathcal{F}_{\alpha\beta}=0. 
\end{equation}
Then we get
\begin{equation}\label{ndet}
\det (\mathcal{G}_{\mu\nu}+(2\pi\alpha')\mathcal{F}_{\mu\nu})
=\det \left(\begin{array}{cc} \mathcal{G}_{\alpha\beta}+(2\pi\alpha')
\mathcal{F}_{\alpha\beta} & 0 \\
0 & 1 \\ \end{array}\right)
=\det (\mathcal{G}_{\alpha\beta}+(2\pi\alpha')\mathcal{F}_{\alpha\beta}).
\end{equation}

When we combine $ \frac{dV}{dT}$  with the second term in (\ref{kt1}),
 we get the equation
\begin{equation}\label{eqT}
T''=\frac{1}{2\pi\alpha'}\frac{dV}{dT},
\end{equation}
where $ T'=\frac{dT}{dx}$. This equation has been solved in
many textbooks (see, for example
\cite{weinberg,kaku}) and we will follow their approach.
 The integration of the previous equation
leads to
\begin{equation}\label{fo}
\frac{dT}{\sqrt{V}}=\sqrt{\frac{2}{2\pi\alpha'}}dx.
\end{equation}
This equation can be easily integrated for the potential given
in (\ref{zeropot}) and we get
\begin{equation}\label{Tsol}
T=T_0\tanh (\frac{m}{\sqrt{2}}x).
\end{equation}

With using (\ref{ndet}), (\ref{fo}) the action (\ref{actlinear}) has a form
\begin{equation}
S=-C_p\int d^p\sigma dx \sqrt{-\det(
\mathcal{G}_{\alpha\beta}+
2\pi\alpha'\mathcal{F}_{\alpha\beta})}2V(T(x)),
\end{equation}
where we have used $V(T)+\frac{2\pi\alpha'}{2}T'^2=2V(T)$.
We can easily integrate over $x$ coordinate
with using (\ref{FGcon}) and we get the final result
\begin{equation}\label{Sfinal}
S=-T_{(p-1)}\int d^p\sigma
\sqrt{-\det (\mathcal{G}_{\alpha\beta}+
2\pi\alpha'\mathcal{F}_{\alpha\beta})},
\end{equation}
where the tension for D(p-1)-brane has a form
\begin{eqnarray}
T_{(p-1)}=C_p\int_{-\infty}^{\infty}dx 2V(T(x))=
C_p \cdot
2V_k\int_{-\infty}^{\infty}dx(1-\tanh^2(\frac{m}{\sqrt{2}}x))^2=\nonumber
\\ 
=\frac{2\pi}{g(4\pi^2\alpha')^{(p+1)/2}}\cdot
\left(\frac{8\sqrt{2}V_k}{3\pi}\right)(4\pi^2\alpha')^{1/2},\nonumber \\
\end{eqnarray}
where $V_k=\frac{(2\pi\alpha'm^2)^2}{16\lambda}$.
As was shown in \cite{SenT}, the vacuum value of the tachyon potential in
the zeroth order approximation 
cancels about $0.60$ of the tension of the non-BPS D-brane, so we have
the value of $V_k $ equal to $V_k=0.60$ and the previous equation gives
the result
\begin{equation}
T_{(p-1)}=0.72\frac{2\pi}{g(4\pi^2\alpha')^{p/2}}
,\end{equation}
which is in agreement with the result \cite{SenT}. We believe that the higher
order correction to the potential as well as using the direct form of the action
(without restriction to the linear approximation)
 (\ref{Naction}) could give a correct value of the tension of a D(p-1)-brane.

We must also stress that we do not obtain any constraints on
 the fermionic degrees of freedom. This follows from the fact that there are no
interaction terms relating left-handed and right-handed spinors with the tachyon
field, since these terms are not allowed through principles of
 T-duality covariance. 
However, this does not contradict  the claim that tachyon condensation on
the world-volume of a non-BPS Dp-brane leads to the action for BPS D(p-1)-brane,
because we must also consider the tachyon condensation in the expression
(\ref{WZ}). It was shown in \cite{Billo} that the tachyon condensation in this
term leads to the correct term for a BPS D(p-1)-brane. With using this
 result and 
(\ref{Sfinal}) the
whole action after tachyon condensation on the world-volume of
a non-BPS Dp-brane has a form
\begin{equation}\label{SSfinal}
S=-T_{p-1}\int d^p\sigma \sqrt{-\det (\mathcal{G}_{\alpha\beta}+
2\pi\alpha'\mathcal{F}_{\alpha\beta})}+\mu_{p-1}\int C \wedge
e^{2\pi\alpha' \mathcal{F}}.
\end{equation}
When we assume that tachyon condensation leads to the correct
values of D-brane tension $T_{p-1}$ and D-brane charge $\mu_{p-1}$, that
(\ref{SSfinal}) is supersymmetric action for D(p-1)-brane 
with $\kappa$-symmetry restored. 
\section{Other proposal for non-BPS D-brane action}
In the recent papers \cite{Berkof1,Garrousi}, the action for
a non-BPS Dp-brane was proposed in the explicit form
\begin{equation}\label{actionBerkofI}
S=-C_p\int d^{p+1}\sigma V(T)
\sqrt{-\det (\mathcal{G}_{\mu\nu}+2\pi\alpha'\mathcal{F}_{\mu\nu}
+2\pi\alpha'\partial_{\mu}T\partial_{\nu}T)}.
\end{equation}
This action is manifestly supersymmetric invariant and also 
T-duality covariant
\cite{Berkof1,Berkof2,Simon2}. This action obeys
the property proposed in \cite{Sen} that for the  tachyon equal to its
vacuum value (this is the value of the tachyon that minimises the
potential $V(T)$) 
$T=T_0, V(T_0)=0$ the action is equal to zero.
 This action incorporates in a very nice way the tachyon
field and the interaction between the tachyon and the massless fields and 
suggests the deep symmetry between the tachyon and the other fields which
was anticipated in \cite{Horava}. 

We would like to discuss the equation of motion for the tachyon
obtained from the linearised form of the action (\ref{actionBerkofI}).
The linearised form of the action is
\begin{equation}\label{actlinearB}
S=-C_p\int d^{p+1}\sigma \sqrt{-\det
(\mathcal{G}_{\mu\nu}+2\pi\alpha'\mathcal{F}_{\mu\nu}
)}\left(\frac{2\pi\alpha'}{2}\tilde{\mathcal{G}}^{\mu\nu}_S
\partial_{\mu}T\partial_{\nu}TV(T)+V(T)\right) \ ,
\end{equation}
from which we obtain the same equation of motion for tachyon
as in (\ref{eq}) with function $F$ now given as
\begin{equation}
F=\frac{2\pi\alpha'}{2}V(T)
\tilde{\mathcal{G}}^{\mu\nu}_S
\partial_{\mu}T\partial_{\nu}T+V(T) \ .
\end{equation}
The analysis of the this equation is the same as in
the previous section and we obtain the same form of the
constrains on the massless fields as before. Much interesting
is the analysis the resulting equation for the tachyon which is
\begin{equation}
\frac{dV}{dT}+\frac{2\pi\alpha'}{2}\frac{dV}{dT}(T')^2-2\pi\alpha'(V T')'=
\frac{dV}{dT}+\frac{2\pi\alpha'}{2}\frac{dV}{dT}(T')^2-2\pi\alpha'(V'T'+VT'')=0,
\end{equation}
where $(...)'=\frac{d}{dx}$. 
We can immediately see that
the solution $T=T_0$ is the solution of equation of motion. This result comes from
the fact that $V(T_0)=0, \left.\frac{dV}{dT}\right|_{T=T_0}=0$ and from the trivial 
fact that the derivative of a
constant function is equal to zero. This confirms the results presented in
\cite{Berkof1}. To obtain the other solution, we multiply the previous
equation with $T'$ and  we get
\begin{equation}
V'+\frac{2\pi\alpha'}{2}V'(T')^2-2\pi\alpha'V'(T')^2
-\frac{2\pi\alpha'}{2}V((T')^2)'=V'-\frac{2\pi\alpha'}{2}(V(T')^2)'=0,
\end{equation}
which can be easily integrated with the result
\begin{equation}\label{VT}
V=\frac{2\pi\alpha'}{2}VT'^2+k \ ,
\end{equation}
where $k$ is an integration constant. We determine this constant from
the fact that in order to get the solution with the finite energy, the solution
must approaches the vacuum value at  spatial infinity, where we
have $V(T_0)=0, T' \rightarrow 0$. We immediately see that
$k=0$. Than the next integration gives
\begin{equation}\label{xT}
T=\sqrt{\frac{1}{\pi\alpha'}}x \ ,
\end{equation}
where this solution does not depend on the precise form of the tachyon
potential. We can show that the solution of the equation of the
motion is given in terms of the function
\begin{equation}\label{Bsol}
T=\left\{\begin{array}{ccc} -T_0 \ , x<L & \\
\sqrt{\frac{1}{\pi\alpha'}}x \ , -L< x<L & \\
T_0 \  , x>L & \\
\end{array}\right. \ ,
\end{equation}  
where the parameter $L$ is determined from the condition that
for $x=L$ the tachyon field given (\ref{xT}) is equal to its
vacuum value
\begin{equation}
T_0=\sqrt{\frac{1}{\pi\alpha'}}L 
\Rightarrow L=T_0\sqrt{\pi \alpha'} \ .
\end{equation}
We see that this solution in the zero slope limit $\alpha'\rightarrow 0$
reduces to the piecewise kink solution discussed in \cite{Berkof1} that
depends only on the vacuum value of the tachyon field. 

The next calculation is the same as in the previous section. The tension of
the resulting D-brane is given as
\begin{equation}
T_{p-1}=C_p\int_{-\infty}^{\infty}dx F(T)=
2C_p\int_{-\infty}^{\infty}dx V(T) \ ,
\end{equation}
where we have used (\ref{VT}). Using (\ref{Bsol})
we obtain from the equation given above
\begin{equation}
T_{p-1}=2C_p\int_{-L}^L dx V(T)=
2C_p\sqrt{\pi\alpha'}\int_{-T_0}^{T_0}dTV(T) \ ,
\end{equation}
For the zeroth order approximation to the potential
(\ref{zeropot}) we obtain the result
\begin{equation}
T_{(p-1)}=(\pi\alpha')^{1/2}C_p\frac{32}{15}\lambda T_0^5=
0.25\frac{(4\pi^2\alpha')^{1/2}}{\sqrt{2}}C_p=0.25T^c_{(p-1)} \ .
\end{equation}
We must emphasise again that in the linear
approximation  this solution does not depend
on the exact form of the tachyon potential, is depends only on
the vacuum value of the tachyon field. 
 However, we must stress that from these simple
calculations we cannot determine the exact form of the action
for a non-BPS D-brane. Perhaps more detailed calculations in
the string theory could answer the question how a DBI action for
a non-BPS D-brane looks like.

\section{Conclusion}

In this paper we have proposed the form of the action for
a non-BPS Dp-brane, which is manifestly supersymmetric invariant, T-duality
covariant and in the linear approximation we have obtained through
the tachyon condensation the supersymmetric action for a  D(p-1)-brane with
$\kappa$-symmetry restored.  We have also discussed the tachyon
kink solution obtained as a solution of the equation of motion which
arises from the variation of the linearised action proposed in \cite{Berkof1}.
We have seen the remarkable fact that we can get the solution which does
not depend on the form of the tachyon potential explicitly, it is function
of the tachyon vacuum value only. At present we cannot determine whether
our proposal is the correct one only on the grounds of supersymmetry
invariance and T-duality covariance.  It seams to us that the more detailed 
calculations in string theory could determine the correct form of the DBI action
for a non-BPS D-brane
\footnote{We thank A. A. Tseytlin for discussing this point.}.

It would be interesting to extend the action (\ref{Naction}) to the 
non-Abelian case, following \cite{Garrousi}. This result could have
the direct relation to the classification D-branes in K-theory \cite{witen,
Horava,Olsen,Olsen2}. We have made some progress in this direction in 
\cite{Kluson2,Kluson3}, where we have tried to extend the action
(\ref{actionKluson}) to the non-Abelian case. It would be nice to
see whether the action presented in \cite{Kluson2,Kluson3} could
be modified in order to be related to the non-Abelian extension of the
action (\ref{Naction}). We hope to return to this question in the
future.

{\bf Acknowledgement:} I would like to thank Zden\v{e}k Kopeck\'{y} and 
especially Rikard
von Unge for conversations and critical comments.


\newpage
\begin{thebibliography}{40}
\bibitem{Sen} A. Sen ,\emph {"Supersymmetric World-volume Action for
Non-BPS  D-branes"}, \jhep{10}{1999}{008}, \hepth{9909062}.
\bibitem{SenA} A. Sen, \emph{"Non-BPS states and Branes in String Theory"}, \hepth{9904207}.
\bibitem{Sen2} A. Sen, \emph{"Type-I D-particle and its Interactions"},
\jhep{10}{1998}{021}, \hepth{9809111}.
\bibitem{witen} E. Witten, \emph{"D-Branes and K theory"},
\jhep{12}{1998}{019}, \hepth{9810188}.
\bibitem{Horava} P. Ho\v{r}ava, \emph{"Type IIA D-Branes, K-Theory and Matrix theory"},
\atmp{2}{1999}{1373}, \hepth{9812135}.
\bibitem{Olsen} K. Olsen and R. J. Szabo,\emph{"Brane Descent Relations in K theory"},
\hepth{9904157}.
\bibitem{Olsen2} K. Olsen and R. J. Szabo, \emph{"Constructing D-Branes >From K theory"},
\hepth{9907140}.
\bibitem{Witen2} N. Seiberg and E. Witten, \emph{"String theory and noncommutative geometry"}, 
\jhep{09}{1999}{032}, \hepth{9908142}.
\bibitem{Billo} M. Billo, B. Craps and F. Rosse, 
\emph{"Ramond-Ramond coupling of non-BPS D-branes"},
\jhep{06}{1999}{033}, \hepth{9905157}.
\bibitem{Aganagic} M. Aganagic, C. Popescu and J. M. Schwarz, 
\emph{" Gauge-invariant and gauge-fixed D-brane actions"},
\npb{495}{99}{1997}, \hepth{9612080}.
\bibitem{SenT} N. Berkovits, A. Sen and B. Zwiebach,
\emph{"Tachyon Condensation in Superstring Field Theory"},
\hepth{0002211}.
\bibitem{Berkovits} N. Berkovits,\emph{"The Tachyon Potential in
Open Neveu-Schwarz String Field Theory"}, \hepth{0001084}.
\bibitem{weinberg} S. Weinberg,\emph{"The Quantum Theory of Fields, Vol.2},
Cambridge University Press,1996.
\bibitem{kaku} M. Kaku, \emph{"Quantum Field Theory"},
Oxford University Press,1993.
\bibitem{Kluson} J. Kluso\v{n},\emph{"Remark about non-BPS D-brane
in Type IIA theory"}, \hepth{9909194}.
\bibitem{Berkof1} E. A. Bergshoeff, M. de Roo, T. C. de Wit, E. Eyras
and S. Panda,\emph{"T-Duality and Actions for Non-BPS D-branes"},
\hepth{0003221}.
\bibitem{Berkof2} E. A. Bergshoeff and  M. de Roo,
\emph{"D-branes and T-duality"},
\plb{380}{1996}{265}, \hepth{9603123}.
\bibitem{Garrousi} M. R. Garrousi, \emph{"Tachyon Coupling
on non-BPS D-branes and Dirac-Born-Infeld Action"},
\hepth{0003122}. 
\bibitem{Simon2} K. Kamimura and J. Simon,
\emph{"T-duality and Covariance of Super D-branes"},
\hepth{0003211}. 
\bibitem{Kluson2} J. Kluso\v{n},
\emph{"D-branes from N non-BPS D9-branes in IIA theory"},
\jhep{02}{017}{2000}, \hepth{9910241}.
\bibitem{Kluson3} J. Kluso\v{n},
\emph{"D-branes in Type IIA and Type IIB theories from
tachyon condensation"}, \hepth{0001123}.
\end{thebibliography}
\end{document}